\documentclass[aps,prb,showpacs,twocolumn,floats,epsfig]{revtex4}
\usepackage{amssymb}
\usepackage{amsbsy}
\usepackage{amsmath}
\usepackage{epsfig}

\begin{document}

\title{Bose-Hubbard model on a star lattice}

\author{Sergei V. Isakov $^{(1)}$, K. Sengupta $^{(2)}$ and  Yong Baek Kim
$^{(3,4)}$}

\affiliation{ $^{(1)}$ Institute for Theoretical Physics, ETH
Z\"urich, CH-8093 Z\"urich, Switzerland.\\ $^{(2)}$ Theoretical
Physics Division, Indian Association for the Cultivation of
Sciences, Kolkata-700032, India.\\ $^{(3)}$ Department
of Physics, University of Toronto, Toronto, Ontario M5S 1A7, Canada. \\
$^{(4)}$ School of Physics, Korea Institute for Advanced Study,
Seoul 130-722, Korea.}

\date{\today}

\begin{abstract}

We analyze the Bose-Hubbard model of hardcore bosons with nearest
neighbor hopping and repulsive interactions on a star lattice using
both quantum Monte Carlo simulation and dual vortex theory. We
obtain the phase diagram of this model as a function of the chemical
potential and the relative strength of hopping and interaction. 
In the strong interaction regime, we find that the Mott phases of
the model at $1/2$ and $1/3$ fillings, in contrast to their
counterparts on square, triangular, and Kagome lattices, are either
translationally invariant resonant valence bond (RVB) phases with no
density-wave order or have coexisting density-wave and RVB orders.
We also find that upon increasing the relative strength of hopping
and interaction, the translationally invariant Mott states
undergo direct second order superfluid-insulator quantum phase
transitions. We compute the critical exponents for these transitions
and argue using the dual vortex picture that the transitions, when
approached through the tip of the Mott lobe, belong to the inverted
XY universality class.

\end{abstract}


\maketitle
\section{Introduction}

The study of microscopic models which may lead to exotic quantum
phases has been carried on for a long time in condensed matter
physics. Recently, one such model system, namely, the
two-dimensional (2D) Bose-Hubbard model on a lattice, has received a
great deal of attention. One of the reasons for this renewed
attention is the possibility of experimental realization of such a
model using cold atoms trapped in optical lattices. \cite{bloch1,
kasevitch1} However, the Bose-Hubbard model is theoretically
interesting in its own right. In particular, it has recently been
pointed out that the superfluid-insulator transitions for fractional
Boson filling factors in this model may be of non Landau-Ginzburg
type in the sense that the low energy theory for these transitions
cannot be described in terms of order parameter fields of the phases
in either side of the transitions. \cite{balents1} Instead, as
pointed out in several studies of the model,
\cite{balents1,burkov1,sengupta1} the transition is aptly described
in terms of vortices which are topological excitations of the
superfluid and whose condensation ultimately leads to
destabilization of the superfluid phase in favor of insulating Mott
phases. \cite{balents1} Such a dual vortex theory provides a list of
possible competing Mott phases via general symmetry requirements of
the underlying lattice. \cite{balents1} In particular the geometric
frustration induced by the lattice structure, which plays a key role
in determining the nature of these competing Mott phases, is quite
naturally described by the dual vortex theory.

Another, more direct and quantitative, approach to studying these
Bose-Hubbard models on a lattice has been numerical quantum Monte
Carlo (QMC) simulations. \cite{isakov1,kuklov1,param1} These QMC
studies provide us with quantitatively accurate  phase diagrams of
the model. Further, they also furnish direct information about both
the nature of the phase transition and the correlation functions in
the Mott phase. \cite{isakov1,param1,wessel1} These features, along
with the possibility of accessing much larger system sizes than
possible in exact diagonalization studies, make them the numerical
method of choice for the lattice Bose-Hubbard models. A combination
of the dual vortex theory and QMC simulation has been recently used
to study the phases of Bose Hubbard model on triangular
\cite{param1,burkov1} and Kagome lattices. \cite{isakov1, sengupta1}
As noted in these works, the Bose-Hubbard model can also be mapped
onto a spin-$1/2$ XXZ model~\cite{sengupta1} leading to
interpretation of the obtained results in terms of both bosons and
quantum spins.

\begin{figure}[ht]
\includegraphics[width=2.6in]{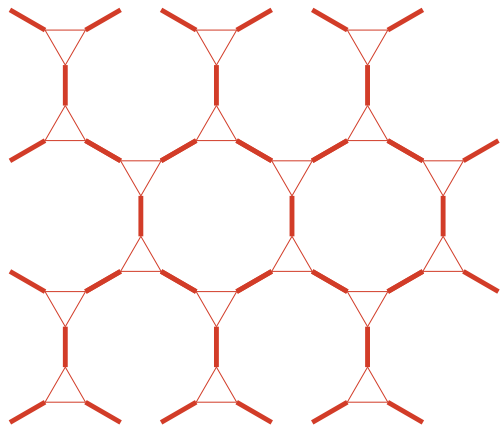}
\caption{The star lattice. The 'triangle' bonds are denoted by thin
lines and the `expanded' bonds are denoted by thick lines.}
\label{lattice1}
\end{figure}

More recently, there have been several studies of the
antiferromagnetic Heisenberg model on a star lattice.
\cite{misguich, richter,review} This lattice, shown in Fig.\
\ref{lattice1}, can be constructed by expanding the sites of a
hexagonal lattice into triangles. It consists of two topologically
inequivalent bonds, triangular and expanded, as shown in Fig.\
\ref{lattice1}. It has been argued in Ref. \onlinecite{richter},
using exact diagonalization studies, that the ground state of the
Heisenberg model on this lattice is a paramagnetic valence bond
crystal (VBC). Such a VBC state is characterized by enhanced
antiferromagnetic correlations along the expanded bonds indicating
singlet formation along those bonds. Such studies have also been
extended for the anisotropic antiferromagnetic Heisenberg model
where the exchange coupling $J^\text{T}$ on triangles is different
from the coupling $J^\text{E}$ on expanded links. \cite{misguich} It
has been argued in Ref.~\onlinecite{misguich} that there may be
another VBC state for $J^\text{T} \gtrsim 1.3 J^\text{E}$ which
consists of $18$-site star pattern. Possible spin liquid and valence
bond crystal phases are also studied in an $Sp(N)$ generalized
model. \cite{choy} These studies, till date, have not been extended
to other spin models such as the XXZ model.

In this work, we study the Bose-Hubbard model, or equivalently, the
XXZ model on a star lattice. The Hamiltonian of the model can be
represented in terms of hardcore bosons as
\begin{eqnarray}
H_b &=& -\sum_{\langle ij \rangle} \left( t^{ij} b_i^{\dagger} b_j +
{\rm h.c}\right ) + \sum_{\langle ij \rangle} V^{ij} n_i n_j -\mu
\sum_i n_i \nonumber\\ \label{bosehubb1}
\end{eqnarray}
where $t^{ij}(V^{ij})$ is the hopping amplitude (interaction
strength) of the Bosons between sites $i$ and $j$, $b_i$ denotes
annihilation operator of the boson at site $i$, $n_i=b_i^{\dagger}
b_i$ is the number density operator for the Bosons at site $i$,
$\langle ij\rangle$ indicates that sites $i$ and $j$ are nearest
neighbors of each other, and $\mu$ is the chemical potential. This
model, in contrast to the Heisenberg model on frustrated lattices,
is amenable to QMC studies. In what follows, we shall allow for
different amplitudes of hopping amplitude and interaction strengths:
$t^{ij}=t^E$ and $V^{ij} = V^E$ for expanded bonds and $t^{ij}=t^T$
and $V^{ij} = V^T$ for triangular bonds. In this work, we shall set
$t^E/t^T = V^E/V^T$. We note at the outset that this model can be
mapped onto an XXZ model in a magnetic field via
a Holstein-Primakoff transformation \cite{sengupta1} and yields
\begin{eqnarray}
H_{\rm XXZ} &=& -\frac{1}{2}\sum_{\langle ij\rangle} J_\perp^{ij}
[S^+_iS^-_j + S^-_iS^+_j] \nonumber\\
&& +\sum_{\langle ij\rangle} J_z^{ij} S^z_iS^z_j + h_z \sum_i S^z_i,
\end{eqnarray}
where $J_{\perp}^{ij}=t^{ij}$, $J_z^{ij}=V^{ij}$, $h_z =
(\mu-1/2)$, and the spin operators are expressed in terms of the boson
operators as $S^+_i=b^{\dagger}_i, S^-_i=b_i, S^z_i=b^{\dagger}_ib_i-1/2$
in the leading order in $1/S$.
We shall use these spin and the boson representations of the model
interchangeably throughout the paper. Note that half filling in boson
language means zero magnetic field in spin language.

The central results reported in this work are the following. First,
using QMC simulation for sufficiently large systems ($L \le 60$) and
low temperatures ($\beta^{-1} \le 0.001J_{\perp}$), we obtain a
phase diagram of the Bose-Hubbard (XXZ) model as a function of the
$\mu/V^E (h_z/J_z^E)$ and $t^E/V^E (J^E_{\perp}/J^E_z)$ for fixed
ratios $V^T/V^E (t^T/t^E)$.  We find that there are three
distinct Mott phases at boson fillings $1/2$ and $1/3$. Second,
using QMC simulations, we compute the equal-time spin-spin
correlation functions and the real space bond-bond correlation
functions for the bosons (spins). From these studies, we
demonstrate that two of these Mott phases, which occur for $J^T=J^E$
at $1/2$ and $1/3$ fillings, are translationally invariant and do
not exhibit density-wave order (magnetization). These Mott states,
in stark contrast to their counterparts in square, triangular or
Kagome lattices, \cite{kuklov1,param1,wessel1} are found to have
resonating valence bonds (RVB) along either the triangle or the
expanded links. The third Mott phase has coexisting density-wave
(N\'eel) and RVB orders and occur for $J^T \ne J^E$ and at $1/2$
filling. These QMC results regarding the nature of the Mott phases
are also supported by qualitative symmetry-based analysis using dual
vortex theory. Third, the translationally invariant Mott states
occuring for $J^T=J^E$ are found to undergo a second-order
superfluid-Mott insulator quantum phase transition with increasing
$t^E/V^E$ (in the spin language, this transition corresponds to a
shift from $S_x$ to $S_z$ ordering if $S_z$ ordering is present or
from $S_x$ ordering to a paramagnet if such an ordering is absent).
This is also in contrast to analogous studies on square, triangular
and Kagome lattices, \cite{kuklov1,param1,wessel1} where QMC
simulations found evidence of either a direct first-order transition
between the superfluid and the Mott phases or an intermediate
supersolid phase. A qualitative symmetry-based analysis using the
dual vortex theory finds this transition to be in the inverted XY
universality class, when the transition point is approached through
the tip of the Mott lobes. This observation is also supported by a
finite-size scaling studies using QMC simulations which yields the
dynamical critical exponent $z$ and the correlation length exponent
$\nu$ for the transition.

The organization for the rest of this work is as follows. In Sec.\
\ref{qmc}, we elucidate our QMC results. We deal with the case
$J_{\perp}^T = J_{\perp}^E$ in Sec.\ \ref{equal1} and follow it up
with the study of the case $J_{\perp}^T \ne J_{\perp}^E$ for
$1/2$ filling in Sec.\ \ref{unequal1}. These numerical results are
then compared with the analytical predictions of the dual vortex
theory in Sec.\ \ref{dvt}. This is followed by conclusion in Sec.\
\ref{conc}.

\section{Quantum Monte Carlo}
\label{qmc}

In this section, we shall analyze the model using quantum Monte
Carlo. We use a multi-site generalization \cite{clustsse} of the
stochastic series expansion (SSE) method. \cite{sse} Here the basic
lattice unit is a site and all its neighbors. Simulations are
performed for systems of linear size $L=12,18,24,36,48,60$ with
$N=6L^2$ sites at different temperatures (the lowest temperature is
$\beta^{-1}=J_\perp/1200$). In Sec.\ \ref{equal1}, we study the
isotropic case with equal hopping and interaction strength on the
triangle and expanded bonds while the anisotropic case is studied in
Sec.\ \ref{unequal1}.

\subsection{$J^\text{T}=J^\text{E}$}
\label{equal1}
\subsubsection{Phase diagram}
\label{pd1}

\begin{figure}[ht]
\includegraphics[width=3.4in]{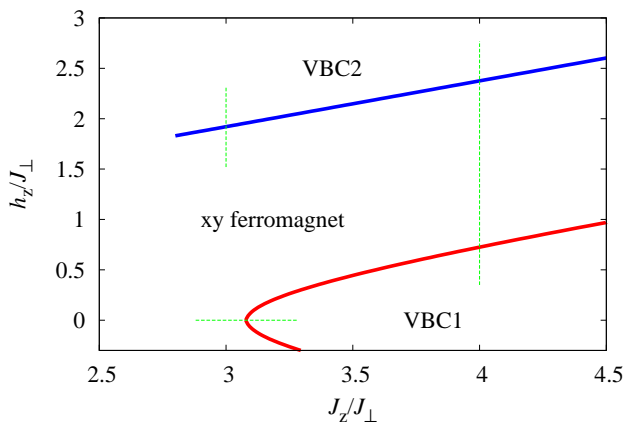}
\caption{The schematic $J^\text{T}=J^\text{E}$ phase diagram from
Monte Carlo simulations. The phase boundaries are denoted by thick
solid lines. Monte Carlo scans are denoted by dashed lines.}
\label{phase:diagram}
\end{figure}

The phase diagram obtained for $J^{T}=J^{E}$ is shown in
Fig.~\ref{phase:diagram}. There are three phases: the XY ferromagnet
and two valence bond crystals denoted by VBC1 and VBC2. In the boson
language, they correspond to the superfluid and Mott phases
respectively. The characteristics of these Mott phases and their
transition to the superfluid phase is discussed in Sec\ \ref{half1}
and \ref{third1} in details. We note here that the Monte Carlo scans
are performed only along three lines shown in the phase diagram so
that the phase boundaries are approximate.

\subsubsection{Half filling}
\label{half1}

In this subsection, we focus on the model in the absence of an
applied magnetic field, {\it i.e.} at $1/2$ filling for bosons.
There is a continuous transition from the superfluid phase to an
insulating phase at $J_z/J_\perp \approx 3.0783$. In the vicinity of
a continuous transition, the spin-stiffness of the XY ferromagnet
(or superfluid density in the boson language), $\rho_s$, which is
measured through winding number fluctuations, scales as
\begin{figure}[ht]
\includegraphics[width=3.4in]{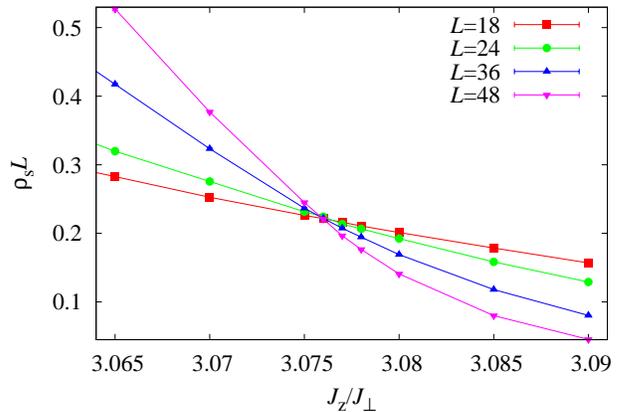}
\caption{Scaling of the superfluid density $\rho_s$ for $z=1$,
$\beta=16L/J_\perp$ and $h_z/J_\perp=0$. Lines guide the eye. In
this and all other figures, error bars are smaller than the symbol
size if not visible.} \label{rhos:scaling}
\end{figure}

\begin{figure}[ht]
\includegraphics[width=3.4in]{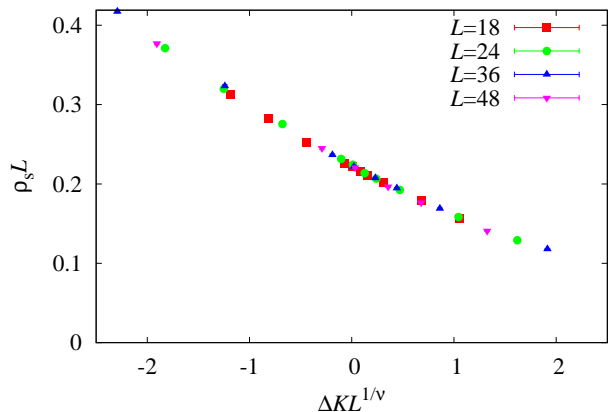}
\caption{Data collapse of the superfluid density $\rho_s$ for $z=1$,
$\nu=0.67$, $(J_z/J_\perp)_c=3.0783$, $\beta=16L/J_\perp$, and
$h_z/J_\perp=0$.} \label{rhos:collapse}
\end{figure}

\begin{equation}
\rho_s=L^{-z} F_{\rho_s}(L^{1/\nu}(K_c-K), \beta/L^z),
\label{eq:rhos:scaling}
\end{equation}
where $F_{\rho_s}$ is the scaling function, $L$ is the linear system
size, $z$ is the dynamical critical exponent, $\nu$ is the
correlation length exponent, $\delta K_c =
K_c-K=(J_z/J_\perp)_c-J_z/J_\perp$ is the distance to the critical
point, $\beta = (k_B T)^{-1}$ is the inverse temperature, and $k_B$
is the Boltzmann constant. It follows from the above finite size
scaling relation (Eq.\ \ref{eq:rhos:scaling}) that the curves for
different systems sizes should cross at the transition point when
$\rho_sL^z$ is plotted as a function of the coupling constant for
$\beta/L^z$ fixed (or for large enough $\beta$ to ensure the ground
state convergence). It also follows from Eq.\ \ref{eq:rhos:scaling}
that the curves for different system sizes should collapse onto a
universal curve for appropriate values of $\nu$ and
$(J_z/J_\perp)_c$ when $\rho_sL^z$ is plotted as a function of
$\delta K_c L^{1/\nu}$. The data scale well with the dynamical
critical exponent $z=1$. In Fig.~\ref{rhos:scaling}, $\rho_s L$ is
shown as a function of the coupling constant. The curves for
different system sizes cross at a distinct point. The data collapse
is shown in Fig.~\ref{rhos:collapse} and leads to a critical
exponent $\nu = 0.67$.

\begin{figure}[t]
\includegraphics[width=1.6in]{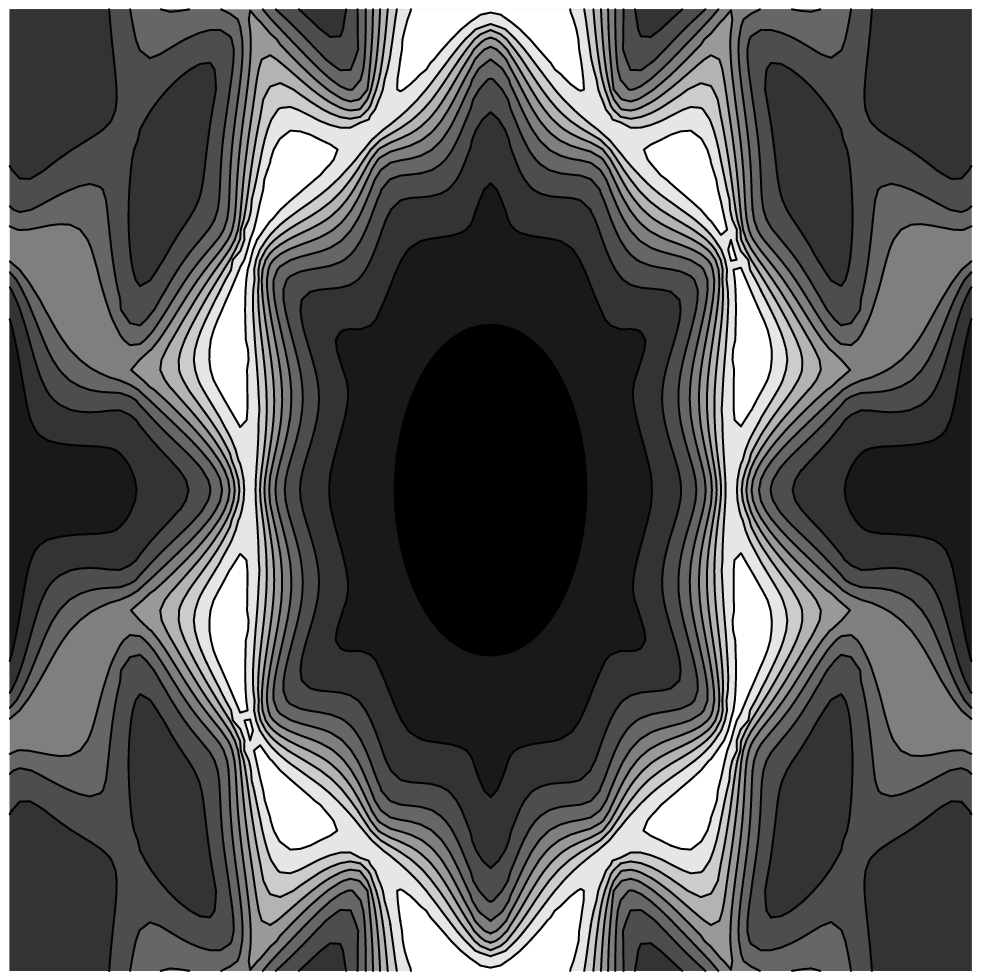}
\includegraphics[width=1.6in]{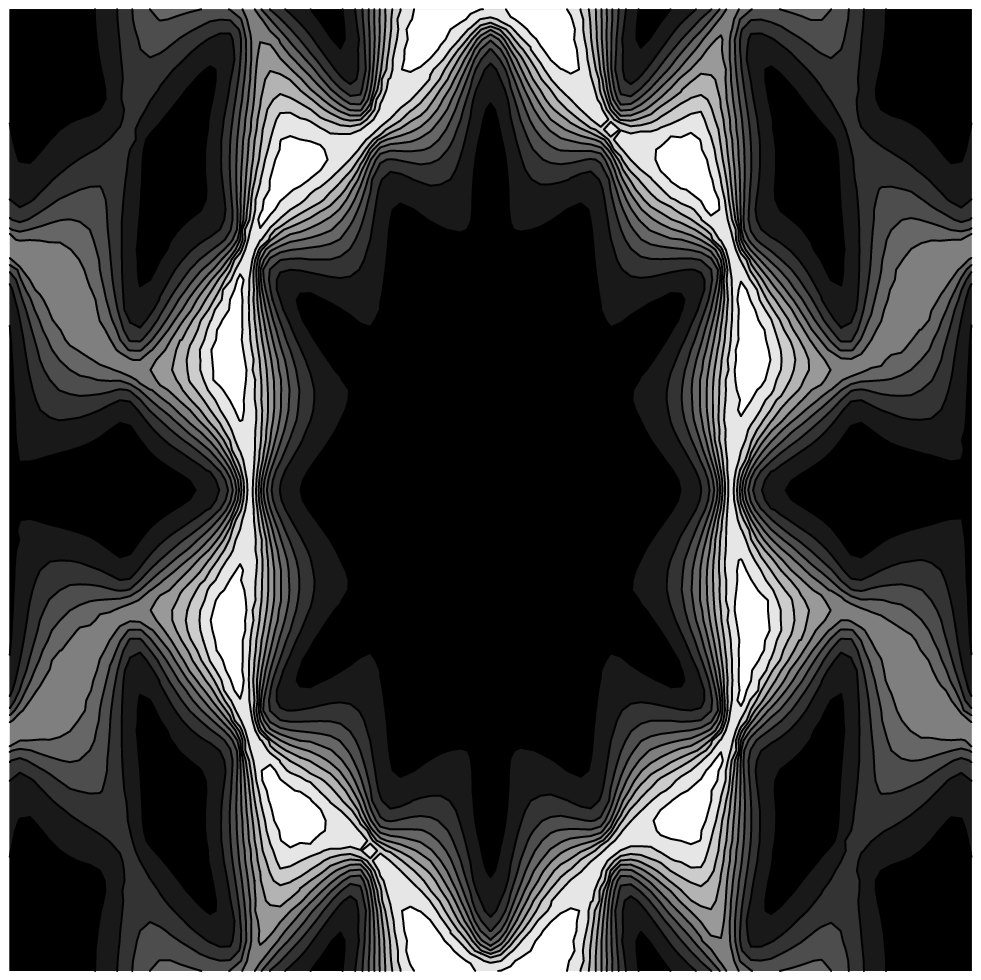}
\caption{The equal time spin-spin structure factor (left panel) and
the static structure factor (right panel) for $L=24$,
$J_z/J_\perp=4$, $h_z/J_\perp=0$, and $k_BT=0.002J_\perp$. The axes
range from $-4\pi$ to $4\pi$.} \label{fig:sf}
\end{figure}

\begin{figure}[t]
\includegraphics[width=3.3in]{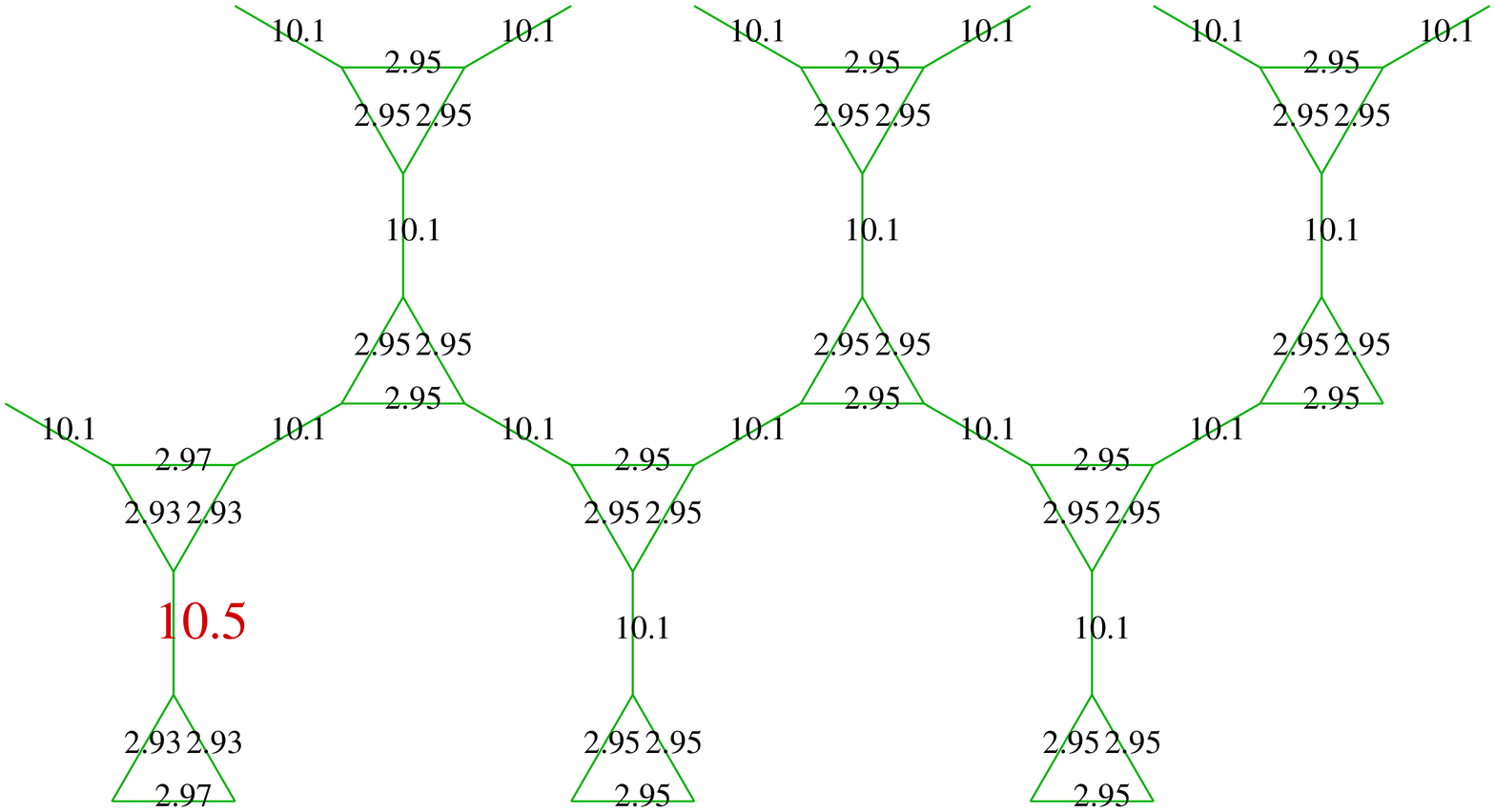}
\caption{ The correlation function $C_{\text b}({\mathbf
r}_1-{\mathbf r}_\delta)$ between the bond indicated by a large text
and the other bonds for $L=24$, $J_z/J_\perp=4$, $h_z/J_\perp=0$,
and $k_BT=0.01J_\perp$. } \label{fig:bcf}
\end{figure}

To address the nature of the insulating phase for $J_z \gg
J_\perp$, we study the equal time and static spin structure factors
that are given by
\begin{eqnarray}
S({\mathbf q})&=&L^2\langle S^{\dagger}_{{\mathbf q}\tau} S_{{\mathbf
q}\tau} \rangle,\quad \chi({\mathbf q})= L^2 \left\langle \int d\tau
S^{\dagger}_{{\mathbf q}\tau} S_{{\mathbf q}0} \right\rangle
\label{sf:chi}
\end{eqnarray}
where $S_{{\mathbf q}\tau}=(1/L^2)\sum_i S^z_{i\tau}\exp(i{\mathbf
q}\cdot {\mathbf r_i})$. The structure factors in the insulating
phase are shown in Fig.~\ref{fig:sf}. There are no sharp peaks which
is a clear indication of absence of magnetic order. The
short-range bow-tie features, found to be present in the structure
factor, are likely to be a remnant of classical dipolar correlations
at finite temperatures, that are known to arise in classical
Heisenberg models on various frustrated lattice.

The nearest neighbor antiferromagnetic spin-spin correlations are
strongly enhanced along the expanded bonds (bonds that connect
triangles) signaling singlet formation along those bonds:
$\langle[S^zS^z]_\text{e}\rangle\approx
2.92\langle[S^zS^z]_\text{t}\rangle$ for $J_z/J_\perp=4$,
$h_z/J_\perp=0$, and $T=0.01J_\perp$. To verify this, we further
compute the real space bond-bond correlation function that is given
by
\begin{equation}
  C_{\text b}({\mathbf r}_\gamma-{\mathbf r}_\delta)=\left\langle
    \frac{1}{\beta}\int B_{\gamma\tau}d\tau
    \int B_{\delta\tau}d\tau \right\rangle,
\end{equation}
where $B_{\alpha(i,j), \tau}=J_\perp(S^x_iS^x_j+S^y_iS^y_j)$ is the
off-diagonal bond operator (at imaginary time $\tau$) of the bond
$\alpha$ connecting spins $i$ and $j$. A plot of $C_b({\mathbf
r}_\gamma-{\mathbf r}_\delta)$ in Fig.~\ref{fig:bcf} shows that the
off-diagonal bond operators are distributed uniformly on the
expanded and triangle bonds with the majority of operators on the
expanded bonds with a ratio $10.1/2.95 \approx 3.42$. There is no
any other bond order. Thus the real-space bond-bond correlation
function also confirms the formation of singlets along the expanded
bonds. This lead us to conclude that the most probable candidate for
this insulating phase is VBC with no symmetry breaking and is
analogous to the VBC state described in Ref.~\onlinecite{richter}.
We denote this phase as VBC1.

\subsubsection{$1/3$ filling}
\label{third1}

\begin{figure}[ht]
\includegraphics[width=3.4in]{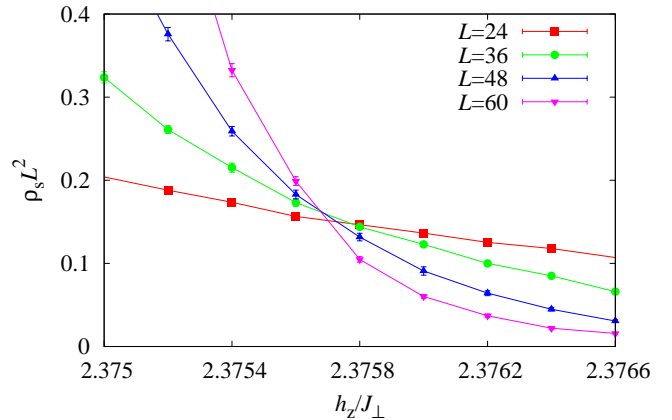}
\caption{Scaling of the superfluid density $\rho_s$ for $z=2$,
$\beta=L^2/3J_\perp$, and $J_z/J_\perp=4$. Lines guide the eye.}
\label{rhos:scaling:2}
\end{figure}

\begin{figure}[ht]
\includegraphics[width=3.4in]{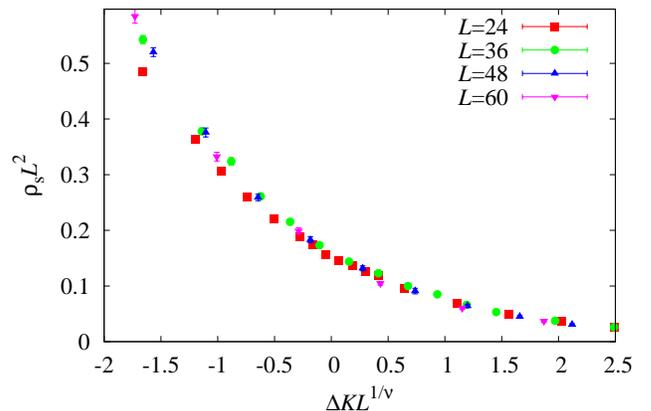}
\caption{Data collapse of the superfluid density $\rho_s$ for $z=2$,
$\nu=0.5$, $(h_z/J_\perp)_c=2.37568$, $\beta=L^2/3J_\perp$. $\Delta
K=(h_z/J_\perp)_c-h_z/J_\perp$ at fixed $J_z/J_\perp=4$.}
\label{rhos:collapse:2}
\end{figure}

\begin{figure}[t]
\includegraphics[width=1.6in]{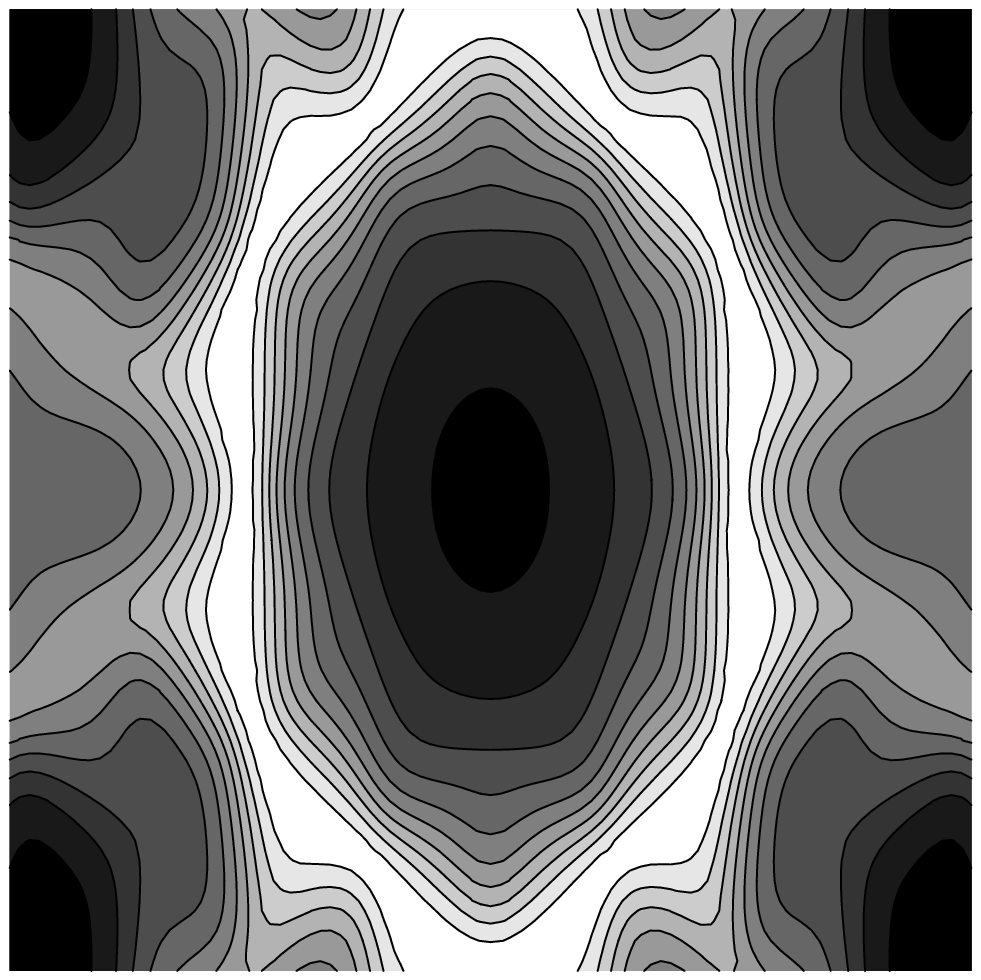}
\includegraphics[width=1.6in]{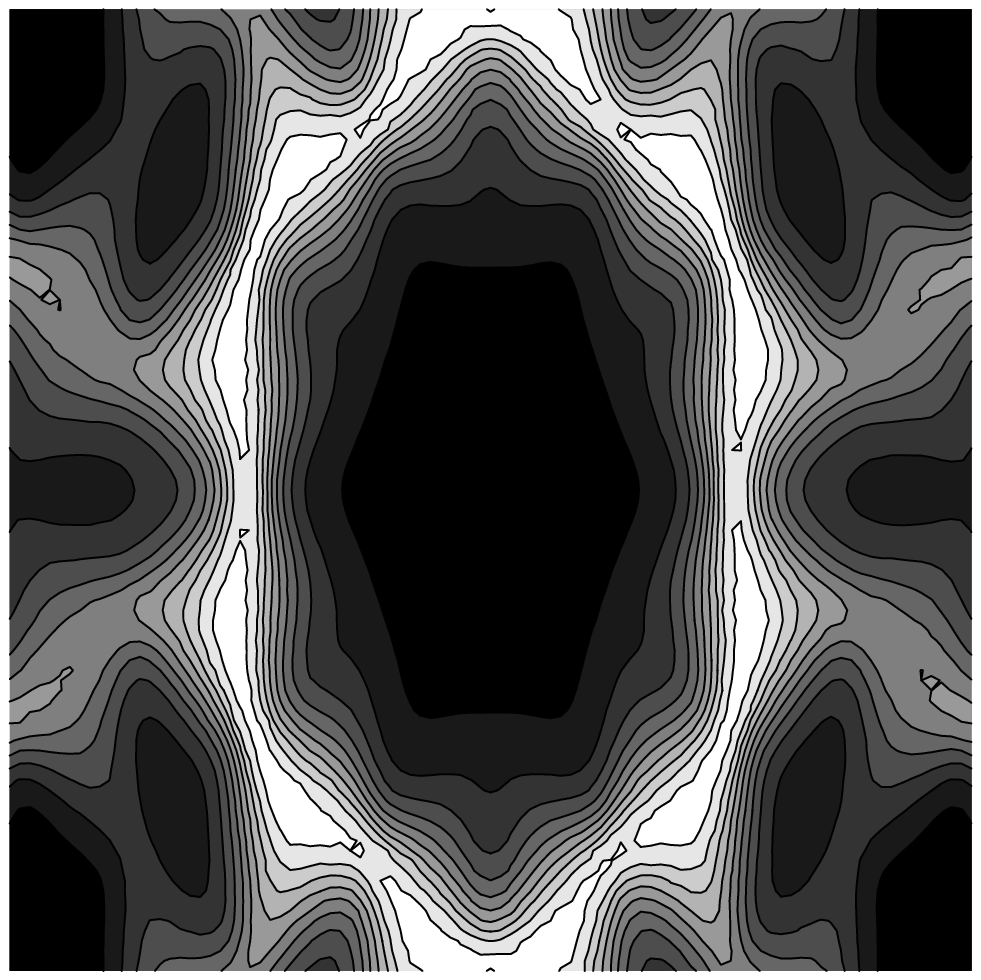}
\caption{The equal time spin-spin structure factor (left panel) and
the static structure factor (right panel) for $L=24$,
$J_z/J_\perp=4$, $h_z/J_\perp=3$ and $k_BT=0.005J_\perp$. Ferromagnetic
peaks due to the uniform background magnetization are subtracted.
The axes range from $-4\pi$ to $4\pi$.} \label{fig:sf:2}
\end{figure}

\begin{figure}[t]
\includegraphics[width=3.3in]{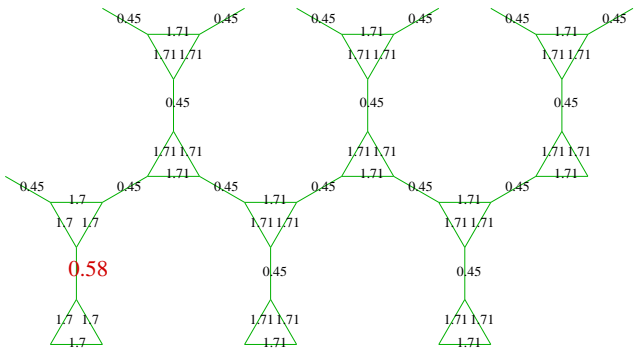}
\caption{ The correlation function $C_{\text b}({\mathbf
r}_1-{\mathbf r}_\delta)$ between the bond indicated by a large text
and the other bonds for $L=24$, $J_z/J_\perp=4$, $h_z/J_\perp=3$,
and $k_BT=0.01J_\perp$. } \label{fig:bcf:2}
\end{figure}

In this subsection, we focus on the phase diagram away from the
$1/2$ filling, or, in the spin language, in the presence of an
applied longitudinal magnetic field. As shown in
Fig.~\ref{phase:diagram}, there is an additional VBC phase (denoted
as VBC2) at the filling of 1/3 (spin magnetization equals $-1/6$).
The transition from the XY ferromagnet phase to the VBC2 phase is
continuous. Within the system size and temperatures that we have
studied, we have not found any signatures of a first order
transition such as double peaked histograms. As shown in
Fig.~\ref{rhos:scaling:2} and Fig.~\ref{rhos:collapse:2}, the data
scale reasonably well with $z=2$ and $\nu=0.5$. It is worth
mentioning that the scaling is not as good as that for the
transition to the VBC1 phase described in the previous section. The
quality of the scaling is probably limited by the fact one needs
data of very high accuracy at extremely low temperatures and very
large system sizes (in order to reach the scaling regime given by
Eq.~\ref{eq:rhos:scaling}) and that is beyond our computational
facilities at the present time.

As shown in Fig.~\ref{fig:sf:2}, the VBC2 phase does not exhibit
magnetic order. The connected spin correlation function is
shorter-ranged than the spin correlation function in the VBC1 phase.
This can be deduced from the larger bow tie width in
Fig.~\ref{fig:sf:2} compared to Fig.~\ref{fig:sf}. The nearest
neighbor antiferromagnetic spin-spin correlations are slightly
weaker along the expanded bonds:
$\langle[S^zS^z]_\text{e}\rangle\approx
0.78\langle[S^zS^z]_\text{t}\rangle$ for $J_z/J_\perp=4$,
$h_z/J_\perp=3$, and $k_B T=0.01J_\perp$. As shown in
Fig.~\ref{fig:bcf:2}, the ratio of the number of the off-diagonal
operators on the expanded bonds to that on the triangle bonds is
$0.45/1.71\approx 0.26$. Thus, in contrast to the VBC1 phase, the
spins resonate along the triangle bonds in the VBC2 phase forming
trimers. There is no other bond order. We can conclude that the VBC2
phase is a quantum paramagnetic phase without any symmetry breaking.

\subsection{$J^\text{T} \ne J^\text{E}$ at $h_z=0$}
\label{unequal1}

\begin{figure}[ht]
\includegraphics[width=3.4in]{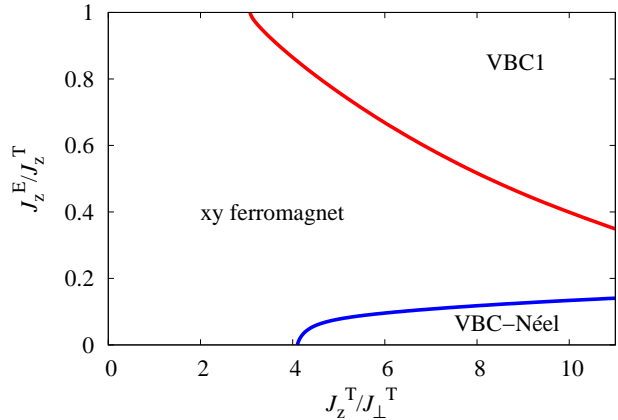}
\caption{The schematic $J^\text{T}\ne J^\text{E}$ phase diagram 
at $1/2$ filling from Monte Carlo simulations. The phase boundaries
are denoted by thick solid lines.} \label{phase:diagram:2}
\end{figure}

\begin{figure}[ht]
\includegraphics[width=3.4in]{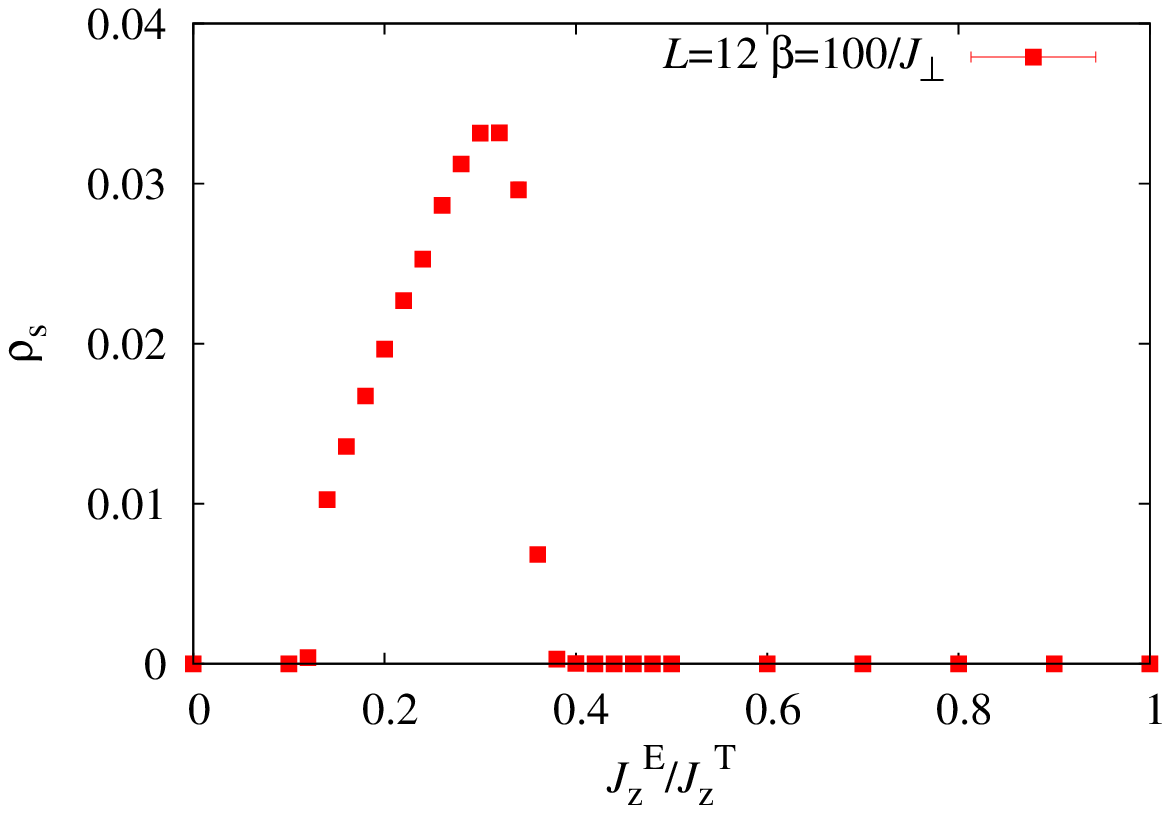}
\caption{The superfluid density $\rho_s$ as a function of
$J_z^\text{E}/J_z^\text{T}$ for $J_z^\text{T}/J_\perp^\text{T}=10$,
$L=12$, and $k_BT=0.01J_\perp^\text{T}$.} \label{rhos:jz10}
\end{figure}

The schematic $J^\text{T} \ne J^\text{E}$ phase diagram at $1/2$
filling is shown in Fig.~\ref{phase:diagram:2}. There are again
three phases: the XY ferromagnet (superfluid in the boson language),
the valence bond crystal phase from Sec.~\ref{equal1} (VBC1), and a
phase that is characterized by both VBC order and N\'eel-like
magnetic order and is denoted as VBC-N\'eel. The latter two phases
are Mott states in the boson language. Monte Carlo scans are
performed only along a few lines so that the phase boundaries are
approximate. We have not attempted to determine the nature of phase
transitions. However, as shown in Fig.~\ref{rhos:jz10}, we find a
narrow region with finite superfluid density between the VBC1 and
VBC-N\'eel phases even for large $J_z^T/J_{\perp}^T$.

\begin{figure}[ht]
\includegraphics[width=3.3in]{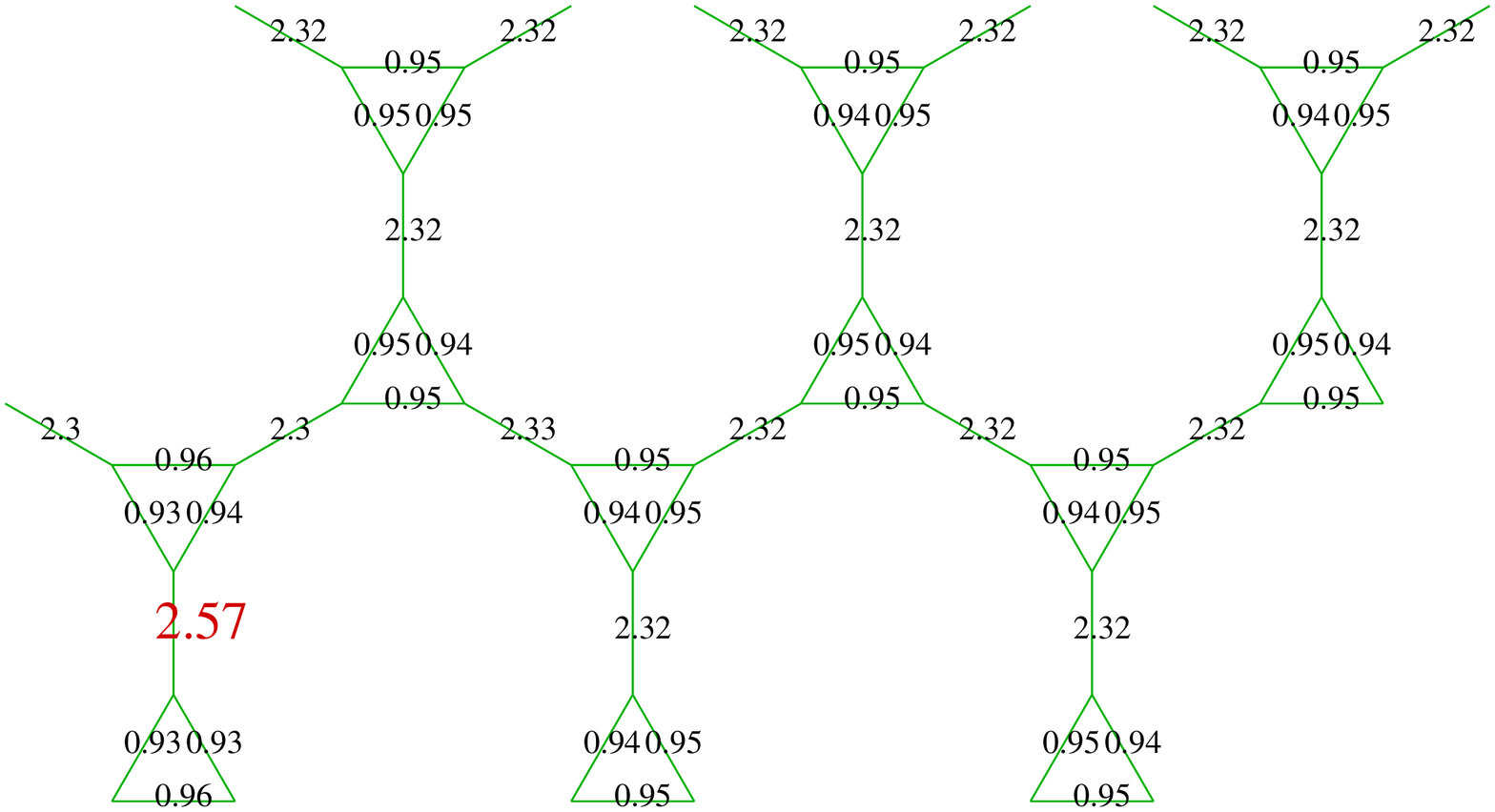}
\includegraphics[width=3.3in]{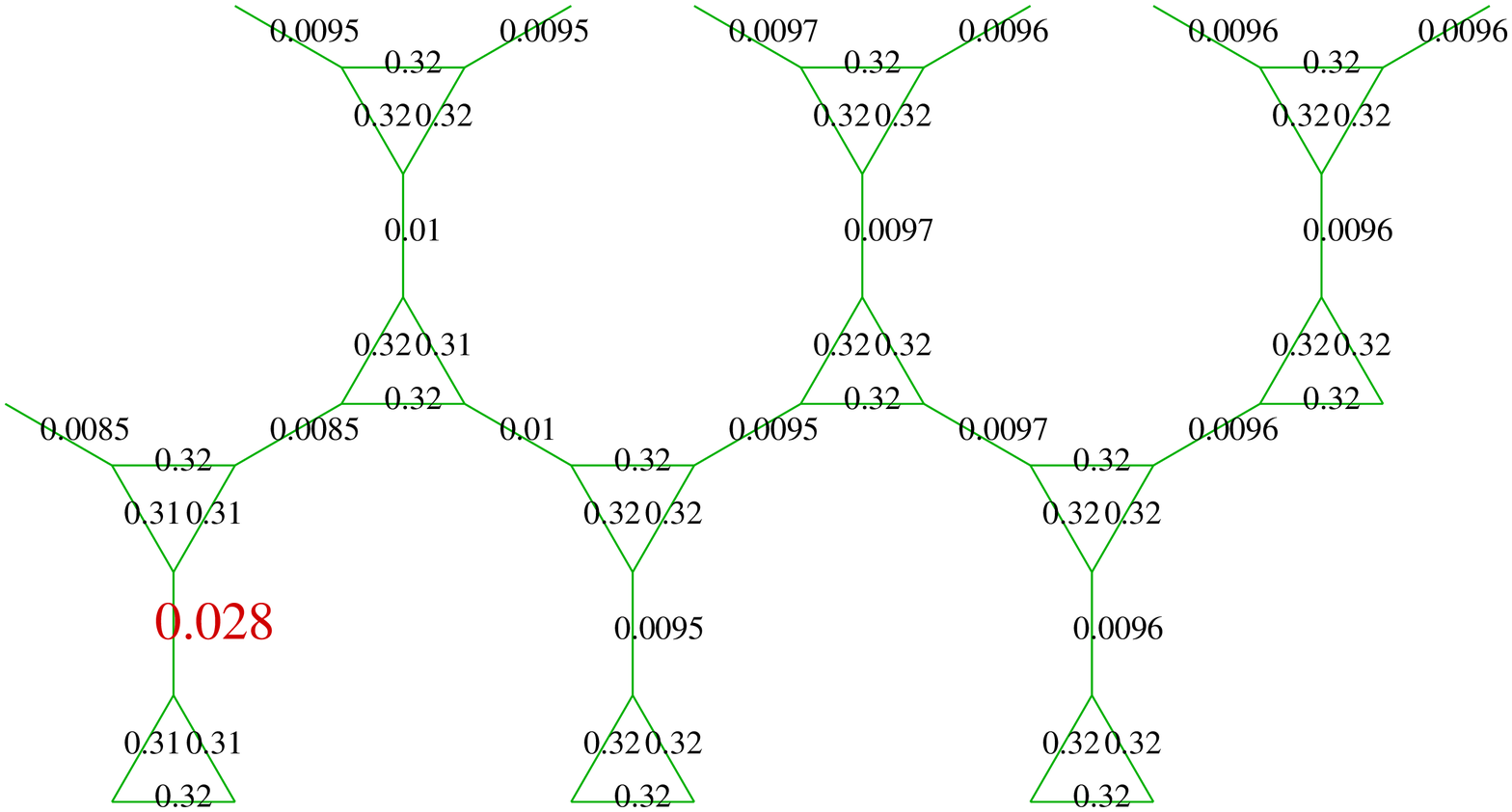}
\caption{ The correlation function $C_{\text b}({\mathbf
r}_1-{\mathbf r}_\delta)$ between the bond indicated by a large text
and the other bonds in the VBC1 phase at $J_z^\text{E}/J_z^\text{T}=0.6$
(upper panel) and in the VBC-N\'eel phase at $J_z^\text{E}/J_z^\text{T}=0.1$
(lower panel) for $J_z^\text{T}/J_\perp^\text{T}=10$, $L=12$, and $k_B
T=0.01J_\perp^\text{T}$.} \label{bcorr:jz10}
\end{figure}

\begin{figure}[ht]
\includegraphics[width=3.4in]{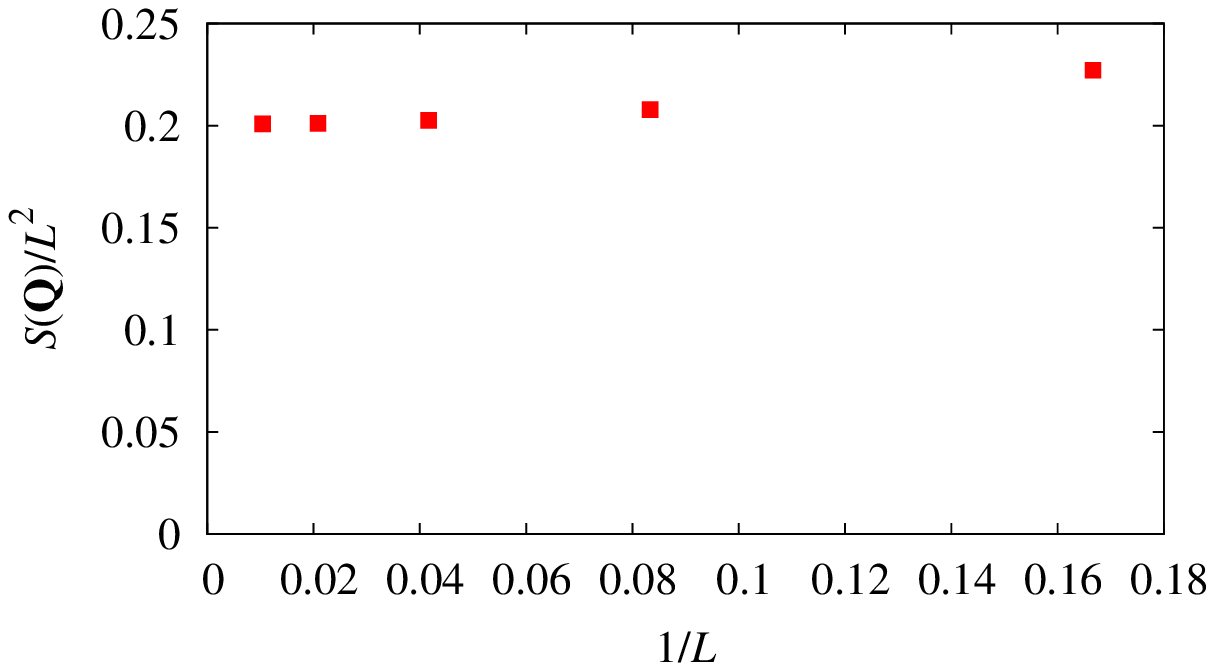}
\caption{The equal time structure factor at the ordering wave vector
$\mathbf{Q}$ as a function of the inverse system size for
$J_z^\text{T}/J_\perp^\text{T}=10$, $L=12$, and
$k_BT=0.01J_\perp^\text{T}$.} \label{sf:jz10}
\end{figure}

In Fig.~\ref{bcorr:jz10}, we show the real space bond-bond
correlation function for different values of
$J_z^\text{E}/J_z^\text{T}$. There are singlets on the expanded
bonds for large $J_z^\text{E}/J_z^\text{T}$ (VBC1 phase) and there
are resonating triangles (trimers) for small
$J_z^\text{E}/J_z^\text{T}$ (VBC-N\'eel phase). The VBC-N\'eel phase
also exhibits long range magnetic order. In Fig.~\ref{sf:jz10}, we
show the finite size scaling of the equal time structure factor
given by Eq.~\ref{sf:chi} at the ordering wave vector ${\mathbf
Q}=(2\pi,2\pi)$. Note that the structure factor vanishes at
${\mathbf Q}=(0,0)$ due to geometrical factors. The structure factor
divided by the number of lattice sites clearly scales to a finite
value in the thermodynamic limit indicating long-ranged magnetic
order. The triangles of the star lattice form the bipartite
hexagonal lattice. The structure of the real space correlations is
such that the spins on two different sublattices of the hexagonal
lattice (belonging to up and down triangles in Fig.~\ref{lattice1})
have antiferromagnetic correlations as in a N\'eel-like ordered
state. This long-ranged magnetic order corresponds to density-wave
order in the boson language.

To understand the VBC-N\'eel state in more detail, consider isolated
triangles ($J^\text{E}=0$). There are two degenerate ground states
per triangle. These are resonating trimers with the total spin
$S^z=-1/2$
$$
  |\psi_1\rangle = \frac{1}{\sqrt{3}}
    (|\mathtt{--+}\rangle + |\mathtt{-+-}\rangle + |\mathtt{+--}\rangle)
$$
and the total spin $S^z=1/2$
$$
  |\psi_2\rangle = \frac{1}{\sqrt{3}}
    (|\mathtt{++-}\rangle + |\mathtt{+-+}\rangle + |\mathtt{-++}\rangle),
$$
where $\mathtt{+}(\mathtt{-})$ indicate sites with $S^z=1/2(-1/2)$.
In the boson language, $|\psi_1\rangle$ and $|\psi_2\rangle$
correspond to one and two bosons per triangle respectively. At $1/2$
filling, the number of spin up triangles is equal to the number of
spin down triangles. For $J^\text{E}=0$, those triangles can be
arranged arbitrary on the star lattice and the ground state is
extensively degenerate. However, finite $J^\text{E}$ selects a
N\'eel state with respect to the total spins of the triangles
because the triangles form a bipartite lattice and the effective
interaction between them is antiferromagnetic. A similar state is
also found in the dual vortex theory analysis, see the next section.


\section{Dual Vortex Theory}
\label{dvt}

In this section, we shall obtain an analytical understanding of the
nature of the Mott phase and the quantum phase transitions from them
to the superfluid phases. Throughout this section, we shall restrict
ourselves to the isotropic case $J^T=J^E$.

The derivation of a dual vortex action starting from the
Bose-Hubbard model (Eq.\ \ref{bosehubb1}) has been elaborated in
Refs.\ \onlinecite{balents1,burkov1,sengupta1}. The vortices are
described in terms bosonic field $\psi_b$ and a dual gauge field
$A_{b\mu}$ which lives on the sites $b$ and links $\mu$ of the dual
lattice respectively. A duality analysis of the Bose-Hubbard model
then leads to an effective dual action which can be expressed in
terms of the vortices and the gauge fields as \cite{balents1}
\begin{eqnarray}
Z &=& \int {\mathcal D} A \int {\mathcal D}\theta
 \exp\left(-S_d\right) \nonumber\\
S_d &=&  \frac{1}{2e^2} \sum_b \left(\epsilon_{\mu \nu \lambda}
\Delta_{\nu} A_{b\lambda} - f\delta_{\mu \tau} \right)^2 \nonumber\\
&& - y_v \sum_b \left(\psi_{b +\mu} e^{2\pi i A_{b \mu}} \psi_b +
{\rm h.c.} \right) \nonumber\\
&&  + \sum_b \left( r \left|\psi_b \right|^2 + u \left|\psi_b
\right|^4 \right) \Bigg]
\end{eqnarray}
where $\psi_b$ are the vortex field living on the site $b$ of the
dual lattice, $A_{b\mu}$ is the U(1) dual gauge field such
that $\epsilon_{\tau \nu \lambda} \Delta_{\nu} A_{b \lambda}= n_i$
where $n_i$ is the physical boson density at site $i$, $\sum_p$
denotes sum over elementary plaquette of the dual lattice,
$\Delta_{\mu}$ denotes lattice derivative along $\mu=x,y,\tau$, and
$f$ is the average boson density. Here $y_v$ is the vortex fugacity
and $r$, $u$, and $e$ denotes parameters of the dual action which
can not be directly mapped onto those of $H_b$ since $S_d$ is not
self-dual to the boson action obtained from $H_{b}$. Therefore we
cannot, in general, obtain a direct mapping between the parameters
of the two actions, except for identifying the magnetic field seen
by the vortices $\epsilon_{\tau \nu \lambda} \Delta_{\nu}
A_{b\lambda}$ as the physical boson density. \cite{balents1,burkov1,
sengupta1} In the remainder of the paper, we shall classify the
phases of this action based on symmetry consideration and within the
saddle point approximation where the gauge fields are pinned to
their saddle point values.
\begin{figure}
\centerline{\psfig{file=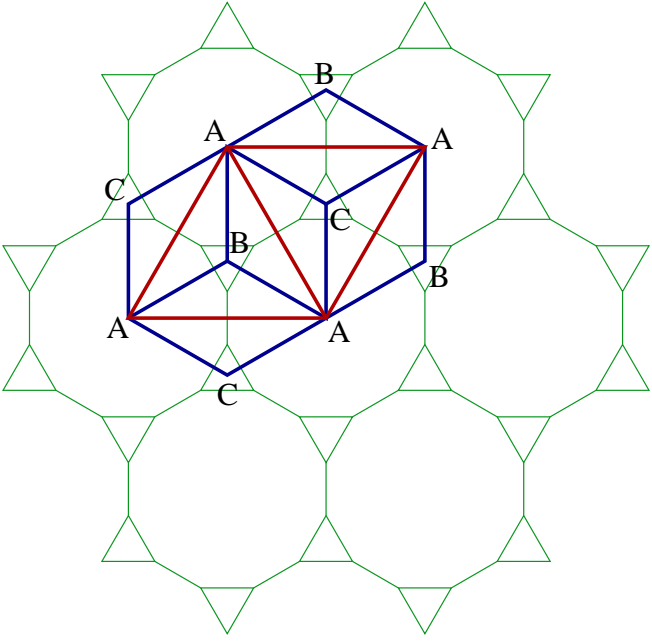,width=\linewidth,angle=0}}
\caption{The star lattice and its dual. Notice that the A sites are
in the center of the hexagon of the dual lattice while the B and the
C sites are in the center of the triangle.} \label{fig1}
\end{figure}
\noindent

The star-lattice and its dual is shown in Fig.\ \ref{fig1}. The
first step towards understanding the phases of $S_d$ within a saddle
point approximation amounts to solving the Hopfstadter problem for
the vortices on this dual lattice shown in Fig.\ \ref{fig1}. We note
from Fig.\ \ref{fig1} that the lattice dual to the star represents a
dice lattice with all diagonals connecting the A sites joined to
each other. The Hamiltonian of the vortices on such a lattice is
given by
\begin{eqnarray}
H = -y_v \sum_{\langle ij \rangle} \sum_{\alpha,\beta= A,B,C}
\left(\psi_{i\alpha}^{\dagger} \psi_{j \beta} e^{ i\gamma_{ij}} + {
\rm h.c.}  \right)
\end{eqnarray}
where $\psi_{i \alpha}\equiv \psi_{\alpha} (a_x,a_y)$ denotes the
annihilation operator for vortex fields at $i \equiv (a_x,a_y)$ and
$\gamma_{ij}$ is the dual magnetic flux, which in the gauge
${\vec A}=H(0,x)$, is given by
\begin{eqnarray}
\gamma_{ij} = 2 \pi f (2a_x/a + \lambda_{\alpha})(y'_j-y'_i)
\end{eqnarray}
where we have used $x_i=a_x$, $x_j= a_x + \lambda_{\alpha} a$, $y'=
2y/(\sqrt{3}a)$, $f = H \sqrt{3}a^2/4\phi_0$ is the flux passing
through an elementary triangle in units of basic flux quanta
$\phi_0$, and $\lambda_{\alpha}$ is a number that depends on
the sublattice index $\alpha = A, B, C$. Here the dual flux passing
through an elementary plaquette is double of that of the dice
lattice. This can be inferred from the fact the present lattice has
two sites of the star lattice in each rhombus (as opposed to one
site of the Kagome lattice in each rhombus) of the dice lattice.
\cite{sengupta1,vidal}

The Schrodinger equation for the vortex fields can be constructed
from the Hamiltonian. With our choice of the gauge, we can write
$\psi_{\alpha}(a_x,a_y) = e^{i\kappa_y y'} \psi_{\alpha} (a_x)$,
where $\kappa_y = 2ak_y/\sqrt{3} \in (0,\pi)$ since we are
restricted within the first Brillouin Zone, and $\alpha = A,B,C$
represents the inequivalent sites of the dual lattice. We thus
obtain, writing energy $\epsilon$ in units of $t$ and defining
$\phi_{\pm}= 2\pi f(2a_x/a \pm 1/2) +\kappa_y$,
\begin{eqnarray}
-\epsilon \psi_B(a_x) &=& \psi_A(a_x+a) + 2 \psi_A(a_x-a/2)
\cos \phi_+ \label{beq} \\
-\epsilon \psi_C(a_x) &=& \psi_A(a_x-a) + 2 \psi_A(a_x+a/2)
\cos \phi_- \label{ceq} \\
-\epsilon \psi_A(a_x) &=& \psi_B(a_x+a) + \psi_C(a_x-a) \nonumber\\
&& + 2 \psi_B(a_x-a/2) \cos(\phi_-) \nonumber\\
&& + 2 \psi_C(a_x+a/2) \cos(\phi_+) \nonumber\\
&& + 2 \psi_A(a_x+3a/2) \cos(2\pi f+ \phi_+) \nonumber\\
&& + 2 \psi_A(a_x-3a/2) \cos(-2\pi f+ \phi_-) \nonumber\\
&& + 2 \psi_A(a_x) \cos(\phi_+ + \phi_-) \label{aeq}
\end{eqnarray}
Note that if we ignore the last three terms in the RHS of Eq.\
\ref{aeq} which involves $\psi_A$, we get back the dual Hopstadter
equation for the dice lattice. \cite{sengupta1,vidal} To solve for
$\epsilon$, we substitute Eqs.\ \ref{beq} and \ref{ceq} in Eq.\
\ref{aeq} and get, using $a_x=3ma/2$ for all A sites,
\begin{eqnarray}
(\epsilon^2-6) \psi_{A m} &=& (4\cos(2\pi f)-2\epsilon) C_m \nonumber\\
C_m &=&   \psi_{A m+1} \cos(6\pi f(m+1/2) + \kappa_y) \nonumber\\
&& + \psi_{A m-1} \cos(6\pi f(m-1/2) + \kappa_y) \nonumber\\
&& + \psi_{A m} \cos(12 \pi f m + 2\kappa_y) \label{eneq}
\end{eqnarray}
\begin{figure}
\centerline{\psfig{file=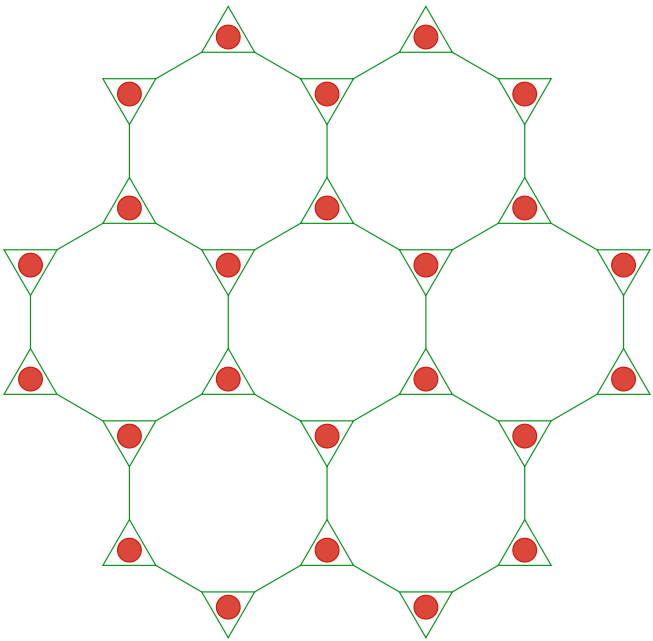,width=\linewidth,angle=0}}
\caption{Mott state at $f=2/3$. The circles indicate a triangle with
either two spin up and one spin down on the three vertices or a
trimerized triangles indicted by circles when superposition of the
spins/bosons due to quantum fluctuations are allowed. The state is
translationally invariant and agrees with the VBC2 state
predicted by QMC when quantum fluctuations are included. The
corresponding ground state for $f=1/3$ can be obtained by simply
flipping the spins. } \label{fig2}
\end{figure}
\noindent
We first consider the case $f=2/3$, which is identical to the
$f=1/3$ filling considered in Sec.\ \ref{third1}. Here Eq.\
\ref{eneq} reduces to
\begin{eqnarray}
(\epsilon^2-6) \psi_{A {\mathbf k}} &=& -2(1+ \epsilon) A_{\mathbf k}
\psi_{A {\mathbf k}}
\nonumber\\
A_{\mathbf k} &=& 2\cos(k'_x)\cos(\kappa_y) + \cos(2\kappa_y)
\end{eqnarray}
where we have taken Fourier transform with respect to $m$ and $k'_x
= 3k_x a/2 \in (0, \pi)$. This has the solution
\begin{eqnarray}
\epsilon_{\pm} &=& -A_{\mathbf k} \pm \left[ \left (A_{\mathbf k}-1\right)^2
+ 5 \right]^{1/2}
\end{eqnarray}
so that the minima of the vortex spectrum occurs at $(k'_x,
\kappa_y) = (0,0), (\pi,\pi)$ and corresponds to $\epsilon = -6$.
Also, substituting the values of $k'_x$ and $\kappa_y$ in Eqs.\
\ref{beq} and \ref{ceq}, we find that $\psi_B(k'_x)=0=\psi_C(k'_x)$
so that the eigenfunctions corresponding to $(k'_x, \kappa_y) =
(0,0), (\pi,\pi)$ are given by
\begin{eqnarray}
\psi_1(m,n) = (1,0,0) \quad \psi_2(m,n) = (1,0,0) e^{i\pi (m+n)}
\end{eqnarray}
where we have used $a_y = \sqrt{3}na/2$ for the A sites. Note that
since $m+n$ is always even for A sites $\psi_1=\psi_2$ and hence the
theory has a single vortex field which should mediate the
transition. Also note that the Mott state is expected to be uniform
since $\Psi_1$ is basically a constant. Thus the simplest state
compatible with these requirement at $2/3$ filling is shown in Fig.\
\ref{fig2}. Here each triangle denoted by a circle, within
mean-field, has two occupied and one empty sites leading to a net
occupancy of $2$ bosons per triangle. Equivalently, in the spin
language, this corresponds to two spin-up and one spin-down sites at
every triangle leading to a net magnetization of $1/2$. However, it
is indeed possible that inclusion of quantum fluctuations will make
the bonds within the triangle to resonate leading to the
trimerized VBC2 state obtained in QMC studies.

The theory of transition pertains to a single vortex field in the
presence of a fluctuating dual gauge field and is thus belongs to
the inverted XY universality class which has $z=1$ and $\nu=2/3$.
\cite{invref} These exponents are the same as their counterparts for
models in the 3D XY universality class. However, the key difference
is that this exponent is obtained for a fully interacting model in
$D=2+1$ in the strong interacting regime. This is in contrast to
$D=3+1$ dimensional systems where one expects the transition to be
fluctuation-driven first order. \cite{halp1} We note that this
expectation, which was initially derived using an $\epsilon = 4-d$
expansion method, is not valid for $D=2+1$ quantum systems where
such transitions remain continuous. \cite{invref} Note that the
quantum phase transition described by this dual vortex theory
requires a fixed density across the transition and hence is valid
when the transition is approached via the tip of the Mott lobe. The
QMC study of Sec.\ \ref{third1} approaches the transition from the
side of the lobe and hence gets a different $z$.

Next we come to case of $f=1/2$. Substituting $f=1/2$, in Eq.\
\ref{eneq}, one gets
\begin{eqnarray}
(\epsilon^2-6) \psi_{A m} &=& -4(1+\epsilon/2) C_m \nonumber\\
C_m &=&   -\psi_{A m+1} \sin(3\pi m + \kappa_y) \nonumber\\
&& + \psi_{A m-1} \sin(3\pi m + \kappa_y) \nonumber\\
&& + \psi_{A m} \cos(2\kappa_y) \label{eneqhalf1}
\end{eqnarray}
Thus here we need to distinguish between the sites which have $m$ as
even and odd integers. Denoting the corresponding fields as $\psi^e$
and $\psi^o$ respectively, we find that
\begin{eqnarray}
\left[\epsilon^2-6 + 4(1+ \epsilon/2) \cos(2\kappa_y)\right]
\psi^e_{A \mathbf k} + C_{\mathbf k}  \psi_{A \mathbf k}^o &=& 0 \nonumber\\
\left[\epsilon^2-6 + 4(1+ \epsilon/2) \cos(2\kappa_y)\right]
\psi^o_{A \mathbf k} + C_{\mathbf k}^{\ast}  \psi_{A \mathbf k}^e &=& 0 \nonumber\\
C_{\mathbf k} = 8 i \sin(k'_x) \sin(\kappa_y) (1+\epsilon/2)
\label{eneqhalf2}
\end{eqnarray}
Note that here $k'_x ,\kappa_y \in (0,\pi/2)$ since the periodicity
in real space has been doubled. From Eq.\ \ref{eneqhalf2}, we find
that $\epsilon = -A_{{\mathbf k} \pm} \pm \sqrt{ \left(A_{{\mathbf k} \pm}
-2\right)^2 +2}$ where
\begin{eqnarray}
A_{{\mathbf k} \pm} &=&  \cos(2\kappa_y) \pm 2 |\sin(k'_x)|
|\sin(\kappa_y)|
\end{eqnarray}
Thus the minima of the spectrum occurs at $(k'_x,\kappa_y) = (
\pi/2, \pi/6)$ with $\epsilon=-3$. Substituting the value of
$\epsilon$ in Eq.\ \ref{eneqhalf2}, we find $\psi^e_A = i \psi_A^o$.
Also, substituting the values of $\psi_A^e/\psi_A^o=i$, $k'_x=\pi/2$
and $\kappa_y = \pi/6$, we find $\psi_B$ and $\psi_c$ from Eq.\
\ref{eneqhalf1}. Finally, this yields the wavefunction
\begin{eqnarray}
\psi_1^{e(o)} &=& [1(-i),0,c(-ci)] e^{i \pi a_1 } e^{i \pi
a_2/6}\label{wavef1}
\end{eqnarray}
where $c=2/3$ , the coordinates of odd and even sites are taken to
be $(a_1,a_2)=(2m+1,2n+1)$ and $(a_1,a_2)=(2m,2n)$ and we haven't
renormalized the wavefunction. Thus, the theory of transition again
pertains to a theory of single vortex field in the presence of
fluctuating dual gauge field and belongs to the inverted XY
universality class with $z=1$ and $\nu=2/3$. This is compatible with
the exponents obtained by QMC study in Sec.\ \ref{half1}.

\begin{figure}
\centerline{\psfig{file=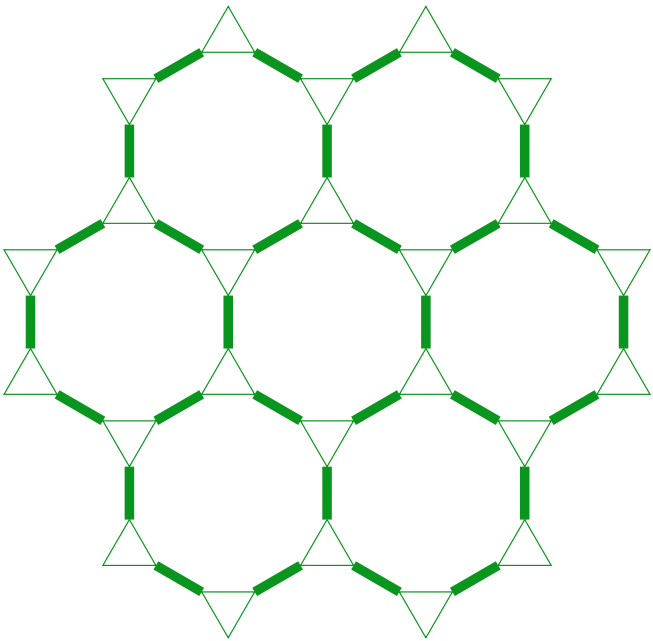,width=\linewidth,angle=0}}
\caption{A possible Mott state with all the connecting bonds of the
triangles forming dimers. The dimerized bonds are shown by thick lines.
This state can occur for $n=0$, $1$, $2$ or $3$ and is 
identical to the VBC1 phase predicted by QMC when quantum
fluctuations are included. See text for details. } \label{fig3}
\end{figure}
\noindent

\begin{figure}
\centerline{\psfig{file=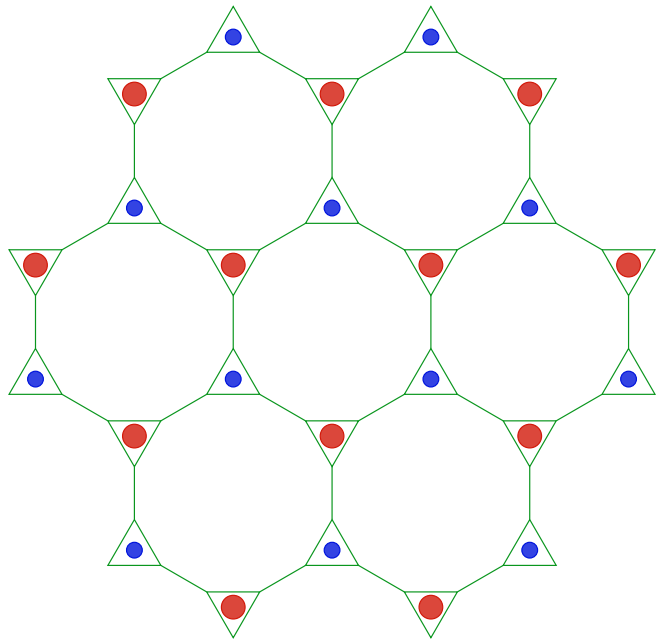,width=\linewidth,angle=0}}
\caption{Another possible state at $f=1/2$ for $n=2$. The large
(red) circles indicate B sites and the corresponding triangles have
one dimerized bond and an up spin while the small (blue) circles
mean a triangle with one dimerized bond and a down spin. A similar
state for $n=1$ can occur and can be obtained by simply
interchanging the sizes (colors) of the B and the C sites. This state
is identical to the VBC-N\'eel phase predicted by QMC when quantum
fluctuations are included. See text for details.} \label{fig4}
\end{figure}
\noindent

Finally, we consider the possible Mott states for $f=1/2$. These are
obtained from qualitative argument outlined below and are shown in
Figs.\ \ref{fig3}, and \ref{fig4}. From the vortex wave function
(Eq.\ \ref{wavef1}), we find that all the B sites of the dual
lattice are equivalent. Thus the triangles of the star lattice whose
centers are occupied by B sites of the dual lattice must have the
same filling and can be filled with $n=0$, $1$, $2$ or $3$ bosons
(or $n$ up and $3-n$ down spins per triangle in the spin language).
This leaves $3-n$ bosons (or $3-n$ up and $n$ down spins) to be
distributed over triangles which contain a C site of the dual
lattice. Such a distribution must have the requisite periodicity of
the wavefunction $i.e.$ a $4 \times 12$ unit cell which involves 4 A
sites (4 hexagons of the real lattice) in the x direction and 12 A
sites (12 hexagons of the real lattice) in the y direction. Thus we
find a multitude of energy-equivalent mean-field states with $4
\times 12$ units cells which corresponds to different ways of
filling these sites keeping the total boson density (or
magnetization) fixed to $3-n$ on the C sites. Similar to the case of
XXZ model on a Kagome lattice analyzed in Ref.\
\onlinecite{sengupta1}, these states can superpose in the presence
of quantum fluctuations leading to a pattern having $n$ Bosons in
every triangle which has a B site of the dual lattice at its center
and $3-n$ bosons (on the average) in every triangles which has a C
site of the dual lattice at its center. Now if we allow the spins on
the joining links of these triangles to hybridize, we get a
translationally invariant dimerized state, shown in Fig.\
\ref{fig3}, where each of the connecting bonds of the triangle can
form a singlet dimer. This state is analogous to the VBC1 state
obtained in QMC study. Another possible state, shown in Fig.\
\ref{fig4} which correspond to $n=2$ or $n=1$, where the one bond in
each of these triangles can hybridize (or form a valence bond),
leads to the translation symmetry broken state shown in Fig.\
\ref{fig4}. This state is analogous to the VBC-N\'eel state found
in QMC studies for $J^T\ne J^E$. There may be other possible states
and a full classification of all of them seems to be difficult. We
point out that the classification of these Mott states necessarily
requires incorporation of quantum-fluctuation induced superposition
between possible mean-field states.

\section{Conclusion}
\label{conc}

In conclusion, we have presented a study of Bose Hubbard or
equivalently spin $1/2$ XXZ model on a star lattice using both QMC
and dual vortex theory. We have shown that for $J^T=J^E$, the
model supports translationally invariant RVB Mott phases at $f=1/2$
and $f=1/3$ and have pointed out that these phases are different
from their counterparts with broken translational symmetry in
square, triangular and Kagome lattices. We have also shown that
these phases, upon increasing the ratio of nearest neighbor hopping
amplitude to interaction strength, undergo a direct second 
order quantum phase transition to a superfluid phase. We have
identified the exponents of this transitions and shown that they
belong to the (2+1)D inverted XY universality class with $z=1$ and
$\nu=2/3$ when approached through the tip of the Mott lobe. When the
transition is approached from the side of the Mott lobe for $f=1/3$,
QMC finds a second order transition with $z=2$. Such clear
signatures of second order quantum phase transitions is in contrast
with the behavior of the model on square, triangular or Kagome
lattice, where these transitions are either first-order or are
accompanied by intermediate supersolid phases. We have also
provided a phase diagram for the system at $1/2$ filling for $J^T
\ne J^E$ and have demonstrated the existence of a Mott phase with
coexisting density-wave (N\'eel) and RVB orders.

\section*{Acknowledgments}

This work was supported by the Swiss National Science Foundation
(SVI); the NSERC of Canada, the Canada Research Chair program, and
the Canadian Institute for Advanced Research (YBK). Simulations were
performed on the Brutus cluster at ETH Z\"urich.

\end{document}